# Effects of Interregional Travels and Vaccination in Infection Spreads Simulated by Lattice of SEIRS Circuits

Yukio Ohsawa*[1], Teruaki Hayashi*[1], Sae Kondo*[2]

*1: The University of Tokyo   *2: Mie University

Email: ohsawa@sys.t.u-tokyo.ac.jp

**Abstract.** The SEIRS model, an extension of the SEIR model for analyzing and predicting the spread of virus infection, was further extended to consider the movement of people across regions. In contrast to previous models that consider the risk of travelers from/to other regions, we consider two factors. First, we consider the movements of susceptible (*S*), exposed (*E*), and recovered (*R*) individuals who may get infected and infect others in the destination region, as well as infected (*I*) individuals. Second, people living in a region and moving from other regions are dealt as separate but interacting groups with respect to their states, S, E, R, or I. This enables us to consider the potential influence of movements before individuals become infected, difficult to detect by testing at the time of immigration, on the spread of infection. In this paper, we show the results of the simulation where individuals travel across regions, which means prefectures here, and the government chooses regions to vaccinate with priority. We first found a weak tendency that vaccination of individuals living in regions of the larger population results in a better performance in suppressing infection spread if one region should be selected. In addition, the active travel of individuals tends to accelerate the spread if the pace of vaccination is as slow as 0.4% of the national population per day, and even a faster vaccination was not sufficient to suppress the wave of spread until May in Japan. However, these findings turned out to be a part of a general law that a quantity of vaccines can be used efficiently by maximizing an index value, the conditional entropy $H_c$, when we distribute vaccines to regions. The efficiency of this strategy, which maximizes $H_c$, was found to outperform that of vaccinating regions with a larger effective regeneration number. This law also explains the surprising result that travel activities across regional borders may suppress the spread if vaccination is processed at a sufficiently high pace, introducing the concept of social muddling.

**Keywords: SEIR model, Regions, Communities, COVID-19, Simulation**

## Main

The spread of COVID-19 since 2019 is now involving all kinds of regions - towns, cities, prefectures, and countries–due to the movements and contacts of people without explicit symptoms. Although testing systems at airports have been introduced and improved, we need further systematic improvements because the virus has spread despite these efforts so far (see the development of governmental policies [1]). It should be noted that the first infection in a country started from a few individuals who immigrated without a positive reaction in the PCR test or any other sign of infection. Thus, borrowing words from the SEIR model [2, 3], attention should be paid to individuals in the infection states as exposed (*E*) or susceptible (*S*) who may change into (*E*) shortly after immigration. Furthermore, we should be aware that people can travel from/to regions such as prefectures or cities within the same country without any tests or even passport control.

Studies have been contributed to the clarification of the influence of travels. For example, using network- or population-based models of infection spread, long-distance movement across regions has been shown to cause a significant increase in the number of infected people [4, 5, 6]. This finding can be related to the lesson *Stay with Your Community* (SWYC) for suppressing infection spread, learned from simulations on a social network model. SWYC means that each individual should avoid meeting as many other unintended people as people to meet intentionally [7] because the excess triggers an explosive spread of infection. The risk of long-distance travel is explained here as the risk of meeting unintended people. On the other hand, we obtain a surprising tendency by combining recent data in [8] and [9], where we find that the increase in the number of travelers and in the new infection cases co-occurred until close to the end of the year 2020; however, their trends started to correlate negatively with each other. That is, people in the USA started to travel frequently since the beginning of 2021, 2 weeks after the start of vaccination, and the effect of the vaccine to suppress new cases obviously overwhelms the expected uptrend due to travels.

Here, we present a lattice of SEIRS circuits where individuals (in a state *S*, *E*, *I*, or *R*) traveling from one region to another are dealt with as different groups but may come into infectious contacts with others in the destination region. We consider the following two strategies in modeling the SEIRS circuit ($S{\rightarrow}E{\rightarrow}I{\rightarrow}R{\rightarrow}S$), explaining the two regions A and B for simplicity.

i) Infected individuals in B may infect others staying in B, including those traveling from other regions.

ii) Individuals traveling from A to other regions are added to the group of the same state traveling from A.

As a result, the model we developed and used for this study is represented in Equations (1) through (4) and illustrated in Fig.2**c**. The details of the logic used to derive the equations are provided in the Methods section. The variables used globally in this paper are listed in Table 1. Borrowing the idea from existing models of vaccination [10, 11, 12], we integrate the doses of vaccination into the reduction of *S*, as in Eq.(1) with $p_{vj}$ as the pace of vaccination in region *j* (the ratio of individuals vaccinated per day in the population of region *j*). Sheer $p_v$ without the suffix *j* denotes the ratio of the number of individuals vaccinated per day to the national population. *v* is the efficiency of the vaccine in reducing the susceptibility. We computed $T_{ij}$ from the data on travelers' flow between regions in 2019 [13]. The lowercase $s_{ik}$, $e_{ik}$, $\psi_{ik}$, and $r_{ik}$ denote the ratio that is division by $N_{ik}$, of the people from region *i* in the states *S*, *E*, *I*, and *R* respectively in region *k*. $\alpha$ is travelers' activity, given by the ratio of the number of travelers relative to before 2020.

Table 1. Variables referred to from multiple sections in this paper.

| | |
|---|---|
| $N_{ij}$ | the number of people originating from region $i$ and staying in $j$ |
| $S_{ij}$ ($E_{ij}$, $I_{ij}$, $R_{ij}$) | the number of susceptible (exposed, infective, or recovered) individuals originating from region $i$ and staying in $j$. $S$, $E$, $I$, and $R$ are called states. The suffixes of states are cut in Fig. 1c for simplicity. |
| $R_t$ ($R_{tj}$) | the effective reproduction number (in region $j$) |
| $T_{jk}$ | The number of travelers from prefecture $j$ to $k$ per day in a normal year free from COVID-19 i.e., 2019 and before |
| $r_1$ ($r_{1i}$) | The percentage of individuals to be infective per day, among exposed ones (who originate from region $j$) |
| $r_2$ ($r_{2j}$) | The percentage of recovering individuals per day, among infective ones (who stay in region $j$). |
| $r_3$ ($r_{3j}$) | The percentage of contacts which cause infections (in region $j$). $r_3 = c R_t$ where $c$ is a constant value. |
| $r_4$ ($r_{4i}$) | the percentage those who return to the susceptible, among recovered ones (who originate from region $i$) |
| $m_I$ | the death rate of infective individuals |
| $m_{SER}$ | the death rate of individuals in $S$, $E$, or $R$ states |
| $p_v$ ($p_{vj}$) | the percentage of vaccine dozes per day in the entire national population (the population of the $j$-th region) |
| $H_c$ | the conditional entropy of vaccine distribution to all regions in Eq.(10) |
| $\alpha$ | traveling activity, that is the number of travels per compared with the normal year (2019), supposed to be uniform in the entire nation |

$$\frac{dS_{ij}}{dt} = \alpha \sum_k (s_{ik} T_{kj} - s_{ij} T_{jk}) + r_{4i} R_{ij} - r_{3j} \frac{\sum_k S_{ij} I_{kj}}{N_j} - v\, p_{vj}\, N_{ij} \quad (1)$$

$$\frac{dE_{ij}}{dt} = \alpha \sum_k (e_{ik} T_{jk} - e_{ij} T_{kj}) - r_{1i} E_{ij} + r_{3j} \frac{\sum_k S_{ij} I_{kj}}{N_j} \quad (2)$$

$$\frac{dI_{ij}}{dt} = \alpha \sum_k (\psi_{ik} T_{jk} - \psi_{ij} T_{kj}) + r_{1i} E_{ij} - (r_{2j} + m_{I_i}) I_{ij} \quad (3)$$

$$\frac{dR_{ij}}{dt} = \alpha \sum_k (r_{ik} T_{jk} - r_{ij} T_{kj}) + r_{2j} I_{ij} - r_{4i} R_{ij} \quad (4)$$

**Fig. 1 (hyperlink)** Two SEIR-based models. **a,** The SEIR model with consideration of the number of infective in-bound travelers ($I_{in}$) in the region receiving travelers [14], corresponding to Eqs (11) through (14) in Methods. **b,** The movement of *S, E, I,* and *R* with the travels, considering the view of the authors and also representing the return of people in *R* to the *S* state as in the SEIRS model [15] due to the loss of once acquired immunity [16] by $r_4$, corresponding to Eqs (15) through (18) in Methods. **c,** The lattice of SEIR circuits: Each vertical alignment i.e., column, of SEIR circuits linked with vertical arrows shows the movements of people originating from the same region, from/to several (three in this figure) regions. In this movement, individuals embrace state *S, E, I,* or *R* changing via interactions as shown by the horizontal arrows. The horizontal alignment represents the interaction of people who originated from various regions and meet in a region.

## Results

### The setting for simulations

We executed simulations considering the virus of variant VOC-202012/01 (lineage B.1.1.7.). Here, $r_1$ was set equal to 0.2 $r_2$ to 0.1, $r_3$ to $0.1 \cdot R_t$ and $m_I$ to 0.012 close to the real death rate of infected cases of COVID-19 in Japan. $r_4$ was set to 0.002 based on reference [16]. $R_t$ was set equal to the value in the real data of the same date in 2020 [17]. On the other hand, according to the increase in the values of $R_t$ for variants according to the literature (an increase of 32% [18], of 43% to 90% [19], etc. according to the literature), $R_t$ is magnified here linearly for the 50 days from April 2021. The value of $R_t$ on the first day 7[th] April was set equal to the value of the same date in 2020, and increased by 40% in 50 days. The vaccine is supposed to reduce the infectivity by 30% by the first dose and 80% by the second dose, to obtain *v* as 55% in Eq. (1). In addition, we estimated $T_{jk}$ using the data in [13]. Note that the point of this paper is to show the general law about the spread of infection and its control using a vaccine, rather than quantitatively correct predictions. The accuracy of simulations as predictions is out of scope. However, we show the results involving B.1.617 originating in India [20, 21, 22] in Supplementary Fig.S1, setting to reach an increase in $R_t$ by 27% from the 51[st] through the 100[th] day of the simulated period, corresponding to a 78% increase in the original virus before B.1.1.7. As shown in this supplementary information, the dependency of the number of infection cases on *Hc* is common to other variants.

The initial values of $S_{ij}$, $E_{ij}$, $I_{ij}$, and $R_{ij}$, as in Eq. (5) through (9), $\Delta t_1$, $\Delta t_2$, $\Delta t_3$, $H_i$, $r_{4i}$, and $\gamma$ are constant values, and all other terms are functions of time *t*. $\Delta t_1$, $\Delta t_2$, and $\Delta t_3$ were set to 2 days, 14 days, and 448 days, respectively. $\gamma$, the number of days to stay when one travels was set to 2.3. infect(*t*) represents the number of newly infected cases on day *t* (data from NHK [23]). The results of the vaccination setting $p_v$ in Eq. (1) to 1%, are shown for the simulated year from April 2021 to March 2022 in Fig. 2, where one prefecture was selected. That is, the number of vaccines was equal (1% of the national population per day) for all cases in Fig 2**a** through 2**f**. In addition, for other paces of vaccination, we found a strong tendency that infections are suppressed when a prefecture of the larger population is selected for vaccination, as in Fig.2**f** and 2**g**, where Pearson's correlation *R* is -0.969 and -0.972 for $p_v$ of 0.3% and 1%, respectively.

However, the number of accumulated infection cases here is larger than $3 \cdot 10^7$, larger than cases hereafter where vaccines are distributed to multiple prefectures.

$$N_{base\ ij}(t) = \gamma T_{ij}(t) \qquad (5)$$

$$E_{ij}(t) = N_{ij}(t)/N_{ii}(t) \text{ avr}_{\tau \text{ in } [t-\Delta t1-2:\ t-2]} \text{ infect}_i(\tau)/r_{1t}, \qquad (6)$$

$$I_{ij}(t) = N_{ij}(t)/N_{ii}(t) \sum_{\tau \text{ in } [t-\Delta t2-2:\ t-2]} \text{infect}_i(\tau), \qquad (7)$$

$$R_{ij}(t) = (1-r_{4i}-) N_{ij}(t)/N_{ii}(t) \sum_{\tau \text{ in } [t-\Delta t3-2:\ t-2]} \text{infect}_i(\tau) \qquad (8)$$

$$S_{ij}(t) = N_{ij}(t) - \{E_{ij}(t)+I_{ij}(t)+R_{ij}(t)\} \qquad (9)$$

**The effect of movements from/to prefectures**

To investigate the effects of travelers' activities, we show cases in which the vaccines are distributed to multiple prefectures. The extent of the diversity of vaccinated prefectures can be represented by the conditional entropy $H_c$, defined in Eq.(10). In Eq.(10), $e_i$ ($i \in \{0, 1\}$) denotes an event that an individual gets vaccinated for $i = 1$ and not for $i = 0$, and $C_j$ ($j \in \{0, 1, ... the\ num.\ of\ regions - 1\}$) means that the individual is vaccinated in the $j$-th region. $p(e_1|C_j), p(e_0|C_j), p(e_1, C_j), p(e_0, C_j)$ are, respectively, equal to $p_{vj}, 1 - p_{vj}, p_{vj} N_{jj}/N, (1 - p_{vj}) N_{jj}/N$. $N$ represents the entire national population.

$$H_c = -\sum_{i,j} p(e_i, C_j) \log p(e_i|C_j) \qquad (10)$$

$H_c$, prevalent in the selection of variables in machine learning (e.g., [24-27]), generally refers to the extent to which vaccines are distributed diversely without a specific intention or causality to choose a region to vaccinate. Each sub-figure in Fig.3 shows a simulated sequence setting a pair of values ($p_v$, $\alpha$) and a distribution of all the vaccines (i.e., percentage of $p_v$ in the national population does in total per day, separated to the 47 prefectures). Because the volume of infection spread depends negatively on $H_c$, as shown later, let us show the values of $H_c$ of the cases depicted in Fig.3**e** through 3**l** among all the 100 simulated distributions for one value pair ($p_v$, $\alpha$), as shown in Fig.3**m** and **n**. A roughly observed tendency here is that frequent travel of people enhances infections in the low range of $p_v$, but this tendency is not obvious for the larger of $p_v$. This change in the tendency with the increase in $p_v$ is one of the essential findings in this paper, which is discussed in the Discussion section.

**Fig. 2 (hyperlink)** The results of vaccinating none or a selected prefecture. **a,** no vaccination, **b,** vaccinating 0.3% of the national population per day i.e., $p_v = 0.3\%$, choosing only people in Tokyo (population $1.4 \times 10^7$, annual average of 2020 $R_t = 1.45$), **c,** Tochigi ($1.9 \times 10^6$, 1.46), **d,** Hyogo ($5.4 \times 10^6$, 1.42), and **e,** Fukuoka ($5.1 \times 10^6$, 1.75). $\alpha$ was set to 40% that is realistic in 2021 according to the data of May 2021. As a result, the number of accumulated infected cases (vertical) is negatively correlated with the population of vaccinated prefectures (horizontal) as shown in **f** ($p_v = 0.3\%$) and **g** ($p_v = 0.1\%$). The Pearson's correlation of both **f** and **g** was 0.97 although the vaccination of multiple prefectures is more effective than **b** as shown later. The subfigures **a** and **b** are also explained in Methods.

**The effect of conditional entropy**

In Fig. 3, we compare the sequences for various values of $H_c$. In this comparison, we set the total quantity of vaccines used to be equal in different sequences. However, the results showed a significant difference in both the peaks and accumulation of the number of infected cases. That is, the larger the value of $H_c$, the more efficiently the vaccination results in suppression of the spread.

**Fig. 3 (hyperlink) a, b, c, d,** newly infected cases per day for different activities of travelers without vaccination. **e, f, g, h,** the results for activities of travelers for $p_v$ = 0.4% of the national population per day. These four sequences correspond to the largest $H_c$ as in (a) of **m**. **i, j, k, l,** the results for activities of travelers for $p_v$ = 1% of the national population per day. These four sequences correspond to the middle valued $H_c$ as pointed by the red arrow in **n**. **m** and **n** show 200 (100 in the upper, 100 in the lower subfigure) of the simulated sequences, represented by 200 dots. (a) (b), (c), (d), (e) in **m** correspond respectively to (a) (b), (c), (d), (e) in Fig.5 i.e., so **f** here is magnified in Fig.5 **a**.

## Discussions

**The overall observations of the results**

As in the comparison between vaccinating people only in Tokyo and in Tochigi in Fig. 2**b** and 2**d**, choosing a region of the larger population causes a stronger suppression of the spread of infection in regions other than the vaccinated one (see the red arrows). Fig.2**f** and 2**g** clarify this tendency, where all the 47 sequences, vaccinating one selected (of the 47) prefecture for one sequence, are plotted. This tendency may sound counter-intuitive because the number of vaccinated individuals is equal in all simulated cases. However, the spread of infection in vaccinated prefectures was substantially suppressed, as shown in the figures. In this sense, this effect of suppressing the spread of infection in the vaccinated local region can be explained by trivial reasoning.

On the other hand, in Fig. 3, we find a significant effect in suppressing the spread of infection by accelerating the vaccination pace $p_v$. From the comparison of 3**a** through 3**d**, 3**e** through 3**h**, and 3**i** through 3**l**, we also find that the spread is enhanced (suppressed) substantially with the increase in $\alpha$. Both of these tendencies occur with the convergence of the trends in different prefectures to a similar sequence, that is, the hills in sequences of the smaller (larger) infection spread are enhanced (suppressed) for the larger $\alpha$ in Fig. 3**a** through 3**h** (Fig. 3**i** through 3**l**) as depicted by red arrows. These changes are frequently observed in the latter half of the sequence. We hypothesize on these observations that the social muddling of non-vaccinated (vaccinated) individuals, respectively, by traveling across regions causes a faster (slower) pace of reproduction that corresponds to creating (cutting off) infectious connections between individuals. Here, social muddling means to cause individuals to contact new others through unintended movements, such as traveling to regions embracing societies new to travelers. If the vaccines are distributed to multiple prefectures at a low or moderate value of conditional entropy $H_c$, the social muddling is expected to increase $H_c$ and work to expand the cutting-off effect through the entire network of people in prefectures connected via the lattice of SEIR circuits. This hypothesis is consistent with the result in Fig. 3**i** through 3**l** (which may be surprising in that the frequent travels are found to suppress the infection spread) because vaccinated individuals are spread by travels and cause social muddling, corresponding to the increase in $H_c$, which disturbs infections.

On the other hand, the travels of non-vaccinated or weakly vaccinated individuals, as shown in Fig. 3**a** through 3**h**, spread the virus to foster infections. Thus, the social muddling effects due to traveling are found to depend on $p_v$ in such a way as to enhance infection for the smaller $p_v$ and suppress infection for the larger $p_v$. Regarding a middle-raged $p_v$, observing Fig. 3**e** through 3**h** more carefully, we find a mixture of these two effects, that is, some prefectures find the suppression of the peaks as Osaka for a large $\alpha$, whereas others suffer from the enhancement, as in the case of Okayama. Borrowing from discoveries using social network models, we can also note that social muddling may cause new contacts between individuals (corresponding to a part of the violation of Stay with Your Community measure, i.e., fostering infectious contacts with unintended individuals [7]), which overwhelms the effect of vaccination if $p_v$ is smaller but will be neutralized or overwhelmed by the effect of disturbing infectious contacts if $p_v$ is larger. As a result, we can expect that the effect of vaccination appears as late as after the social muddling of vaccines affects a wider range in the social network. As in Fig. 4, the early hill till May 2021 called the "fourth wave" in Japan had not been suppressed by vaccines even if the $p_v$ was as quick as 1% (the real pace was less than 0.2%) of the population per day. The effect of improving the vaccination pace from 0.1% to 1% per day was found to be substantial in the nationwide view after June.

**Conditional entropy as a key factor for optimizing vaccination strategy**

The above effect of choosing large-population regions in vaccination, as shown in Fig. 2, is not as substantial as that of distributing the vaccine diversely to regions as in Fig. 3, as mentioned above. As shown in Fig. 5, the distribution of vaccines with a larger value of $H_c$ tends to result in a more significant suppression of the spread of infection. If this is a valid tendency, maximizing $H_c$ can be regarded as a policy for minimizing the spread of infection. And to maximize $H_c$, we can mathematically propose (1) and (2) below as useful political strategies of vaccination based on Eq.(10).

(1) If a region should be selected for vaccination, a larger population is recommended.

(2) If more than one region can be considered, a given quantity of vaccine should be distributed without an intentional bias to a restricted region.

To evaluate the tendency above, Fig.6**a** through 6**l** were obtained by varying the vaccination pace $p_v$ from 0 to 1% and $\alpha$ from 0.13 through 2.0, collecting 100 random sample sequences for each condition given by a pair of ($p_v$, $\alpha$). For each condition, the simulation was executed for patterns of distributing the vaccine to regions by randomly setting $p_{vi}$ of each (*i*-th) prefecture, which means the percentage per day of individuals who have received vaccination in the population of the *j*-th region, under the constraint that $\sum_j p_{vj} N_{jj}$ is equal to $p_v \sum_j N_{jj}$ i.e., the total number of vaccines used per day is given by the pace of vaccination relative to the national population. As shown in each subfigure among Fig.6**a** through 6**l**, the increase in $H_c$ negatively affects the number of infection cases by a stronger impact than the sheer choice of a large-population region to vaccinate, as shown in Fig. 2. This effect ranges from moderate to strong negative correlations (Pearson's coefficient *R,* embedded in the subfigures, ranging between close to 0.5, and over 0.7) for $p_v$ of 0.2% and larger, as shown in the subfigures. Furthermore, for the cases with B.1.617 originating in India, as shown in Supplementary Fig.S1, the negative correlation of the number of infection cases on $H_c$ for $p_v > 0.2\%$ is more significant according to the Pearson's correlation values.

It may be inferred that the average $R_t$ of vaccinated regions should be considered as a measure to optimize the distribution of vaccines to regions. This is because $R_{tj}$ of a region *j* determines how quickly and widely the region suffers from infections and spreads them to other regions. According to our simulations, the dependency of the accumulated number of infection cases on $R_t$ representing the average of $R_{tj}$ for all the regions (i.e., *j*'s), weighted by the number of vaccinated individuals, is shown in Fig.6**m** through 6**x** for comparison with Fig.6**a** through 6**l**. By viewing, we find that the dependency on $H_c$ is more significant than $R_t$, except for the case of $p_v = 0.1\%$. Quantitatively, Pearson's correlation values support this intuition. Thus, we can approximately optimize the distribution of vaccines by maximizing $H_c$ for $p_v > 0.1\%$ and, and by maximizing $R_t$ for a smaller $p_v$. It should be noted, however, $R_t$ is not easy to use for this purpose because the average of $R_t$ in the target time range (April 2021 through March 2022 in this paper) is unstable and is not trustworthy for estimating its future value if there are some causes to change people's social activities changes, such as Olympic games. Furthermore, we find a difference in Supplementary Fig.S1 from Fig.6, that is, the positive correlation of the number of infected cases to $R_t$ in the case of B.1.617, which is negative in the case of B.1.1.7. This tendency is because the too fast infection spread in a prefecture of too large $R_t$ cannot be conquered by the slow vaccination, which works more efficiently if used in prefectures with a smaller $R_t$ (smaller, but in the largest range in the case of B.1.1.7) within the range possible to conquer even by the slow vaccination. In other words, such a slow vaccination can help only slowly spread prefectures from B.1.617. Thus, $H_c$ can be regarded as easier to use than $R_t$, considering the stability of regional populations and more generally useful for optimizing the distribution of vaccines considering the stability of negative correlation.

Another noteworthy tendency shown in Fig.6 is the traveling activity. That is, the more frequent travel across the borders of prefectures represented by the larger $\alpha$ causes a significant increase in the number of infected cases below the vaccination pace $p_v$ of 0.001, but the increase is moderate as $p_v$ is improved to close to 0.4% and then decreases if the vaccination pace is further increased ($p_v = 1\%$). As shown in Fig.6**y**, $p_v$ of nearly 0.4% is the borderline of this reversal turn. As shown in Fig. 6**z**, the suppression of infection spread with an increase in $\alpha$ for $p_v > 0.4\%$ is not as significant as the acceleration with the increase in $\alpha$ for $p_v < 0.3\%$. For the cases with B.1.617 originating in India, as shown in Supplementary Fig.S1, the suppression of infection spread with the increase in $\alpha$ was found for $p_v > 0.6\%$, as well as a significant negative correlation with $H_c$ for $p_v > 0.1\%$. We should say "the risk due to travels can be suppressed" rather than "it is encouraged to travel" across prefectures by setting large $p_v$ and $H_c$.

## Conclusions

The first two findings in this paper, that is, (1) the focused vaccination in regions of the larger population as well as of the larger $R_t$ tends to be the more effective for vaccination, and (2) the vaccines hardly work to suppress the infection spread for a month since the start of vaccination (so called the 4th wave of April in Japan), respectively, coincide with the general tendencies shown so far in studies having worked with the Japanese Cabinet Office ([14, 28]). Therefore, we conclude that the lattice of SEIRS circuits shows a reasonable performance as a whole, which partially supports the reliability of this method in estimating the risks in local regions. This enabled the following discoveries of the overall trend of a group of regions, such as a nation, considering the interaction of micro (regional) level and macro (national or global spread) level spread of infections. The new practical findings here are (4) a restricted quantity of vaccine can be used efficiently by maximizing the conditional entropy, (5) the travel across the borders of regions accelerates the infection spread if the vaccine is distributed at a slower pace as $p_v$ <0.003 in the national population per day, but the spread due to travel is substantially suppressed if the pace is improved to $p_v$ >0.4% ($p_v$ >0.6% for B.1.617).

So far, the principle of staying with one's community [7] has been shown to reduce the risk of travel. That is, the spread due to long-distance travel can be calmed as far as individuals other than the travelers stay with their communities [28], in spite of their influence on enhancing the spread of infection, as shown using a social network model [4] and a population-based model [5]. For the time being, mixtures of state-transition models such as SEIR, its various extensions, and network models are on the way to be combined to obtain unified models of interacting microscopic agents and the macroscopic behaviors of the society [29, 30, 31]. In comparison, the method proposed in this paper can be positioned as a method to model the network of societies rather than of individuals to understand mesoscopic social interactions such as social muddling. The authors are designing multiscale spatiotemporal data for modeling human and societal behaviors using methods for data-interactive innovations [32].

**Fig. 4** (hyperlink) The weak impact of vaccination in the first small hill i.e., the vaccine does not suppress the spread for the first one month (see the small hill in the left-most). The subfigures **a**, **b**, and **c** here are respectively the magnification of Fig.3**b**, 3**f**, and 3**j**.

**Fig. 5.** (hyperlink) The sequences of daily infected cases for varied values of conditional entropy $H_c$ for a fixed total vaccination $p_v = 0.4\%$/day and a constant travel activity $\alpha = 0.4$. The sequences correspond to the arrows in Fig.3**m**.

**Fig. 6** (hyperlink) The effect of conditional entropy $H_c$ (**a-l**) and the effective reproduction number (**m-x**) on the number of infection cases (accumulation), varying the vaccination pace $p_v$ and the travel activity $\alpha$. **y**, the effect of vaccination pace $p_v$ on the average number of infection cases (accumulation) for $\alpha$ of 0.13, 0.3, 1.0, and 2.0. Standard deviations or confidence intervals are put out of scope of **y** but used in **z** where the p-value as a result of t-test is shown as an index of significance of the effect of $\alpha$ on the number of infection cases for each value of $p_v$ (p-value is usually used for checking the significance by comparing with a borderline value e.g. "p < 0.05"). The number of infection cases is significantly positively dependent on $\alpha$ for the range of $p_v$ <0.3%, negatively dependent on $\alpha$ for the range of $p_v$ >0.4%.

## Methods

### The SEIR model and its Extensions

The basic idea of the SEIR model originated from the SIR [33-35], where *S*, *I*, and *R* refer to the following numbers of people:

*S*: Number of *susceptible* individuals When a susceptible and an infectious individual come into contact with a risk of infection (e.g., 15 min within a distance of 2m), the susceptible may catch the virus and transits to *I* below.

*E*: Individuals who have been exposed but are in an incubation period, during which one may have caught the virus but is not yet infective.

*I*: The number of *i*nfective individuals who have been infected and may infect individuals in *S*.

*R*: Number of individuals who had once infected and then *recovered*. Some analysts deal with dead individuals as a part of *R* (called *removed* in such a case), but below we count the dead as a part of *I* who do not transit to *R*.

SIR has been revised to SEIR by adding *E* below because the period of incubation, during which the individual may move from state *S* to *I*, was not negligible.

To include the influence of travelers from other regions, the number of infective people has been previously represented by $I_{in}$, which means the in-bound risk in the region receiving travelers [13]. Thus, Equations (11) through (19) were used in the analysis and the simulation of infection spread. See Fig.1**a** for an illustration of these equations.

$$\frac{dS}{dt} = m_{SER}(N - S) - r_3 S (I + I_{in})/N \qquad (11)$$

$$\frac{dE}{dt} = -(r_1 + m_{SER}) E + r_3 S (I + I_{in})/N \qquad (12)$$

$$\frac{dI}{dt} = -(r_2 + m_I) I + r_1 E \qquad (13)$$

$$\frac{dR}{dt} = -m_{SER} R + r_2 I \qquad (14)$$

### Adding the effects of in-bound travelers to SEIR model

Considering the above view of the authors and also representing the return of people in *R* to the *S* state as in the SEIRS model [14] due to the loss of once acquired immunity [15] by $r_4$, let us next consider Equations (15) through (18), as shown in Figure 2(b). $S_{in}$, $E_{in}$, $I_{in}$, and $R_{in}$ refer to the in-bound travelers coming to be merged in the target region with others in the states *S*, *E*, *I*, or *R*.

$$\frac{dS}{dt} = S_{in} + r_4 R + m_{SER}(N - S) - r_3 S I/N \qquad (15)$$

$$\frac{dE}{dt} = E_{in} - (r_1 + m_{SER}) E + r_3 S I/N \qquad (16)$$

$$\frac{dI}{dt} = I_{in} - (r_2 + m_I) I + r_1 E \qquad (17)$$

$$\frac{dR}{dt} = R_{in} - (r_4 + m_{SER}) R + r_2 I \qquad (18)$$

However, we should be careful that movements are mutual interactions between multiple regions rather than a one-way transition of an individual. For example, imagine a world where only two regions, A and B, exist. Ten thousand people from region A travel to region B for a few days. These people return to region A after the travel, so the change in the number of individuals in all sections $S_{in}$, $E_{in}$, $I_{in}$, and $R_{in}$ depends on the difference between the densities of $S$, $E$, $I$, and $R$ in regions A and B if the same number of people going out to region B returns to region A. In other words, $X_{Bin}$ is estimated using Eq.(19) for state $X$ as one of $S$, $E$, $I$, and $R$ (we assume $N_{Ain}=N_{Bin}$, i.e., all people return to the departure region).

$$X_{Bin} = \left(\frac{X_A}{N_A} - \frac{X_B}{N_B}\right) T_{AB} \qquad (19)$$

Here, $T_{AB}$ represents the number of travelers from region A to B. $X_{Bin}$ reflects the difference between the densities of the population in state $X$ in regions A and B, on the estimation of the incoming volume to B. That is, $X_B/N_B$ (or $X_A/N_A$) refers to the density of state $X$ among all who return to region A from B (or go out to region B from A). Because it is mostly just a few days (2.3 days in average in Japan [16]) between the time travelers go from region A to B and the time they return to A, during which the ratio of the travelers of state $X$ changes by an incomparably smaller value than $X_A/N_A$, $X_B/N_B$ is expected to be close to $X_A/N_A$. This point can be expressed as follows:

The number of travelers from region A to region B: $\rho N_A$

Number of travelers in state $X$ from region A to region B: $\rho X_A$

Travelers returning to region A from region B: $(1+\delta_1) \rho N_A$

Travelers returning to region A from region B in state $X$: $(1+\delta_2) \rho X_A$

Here, $\rho$, $\delta_1$, and $\delta_2$ denote the ratio of travelers from A to B among all people in A, the ratio of those who are added (can be less than zero because some people may choose to stay longer in B) on the way back to A, and the ratio of those who get newly (minus those who recover) infected within a few days of travel if $X$ is $I$. The value in place of $X_B/N_B$ in Eq.(19) is thus $(1+\delta_2)/(1+\delta_1) (X_A/N_A)$, that is, the ratio of those in state $X$ among the returning travelers, is nearly equal to $X_A/N_A$ because $|\delta_1|<<1$ and $|\delta_2|<<1$, considering the short travel time of as 2.3 days. However, if regions A and B have a significant difference, for example, the infection in B is substantially more serious than A, $X_B/N_B$ comes to be computed as significantly larger than $X_A/N_A$. Owing to this gap, the equation set (Equations (15) through (19) or Fig.1**b**) cannot fit the problem.

**Lattice of SEIRS Circuits**

To cope with this problem, we propose a framework in which individuals (in a state $S$, $E$, $I$, or $R$) originating from region A and traveling to B are dealt with as different groups but may come into infectious contacts with others in B. When they return from region B back to region A, their probability of being in state $X$ should be estimated as close to $X_A/N_A$. instead of $X_B/N_B$. Thus, we consider the following two strategies in modeling the SEIRS ($S \rightarrow E \rightarrow I \rightarrow R \rightarrow S$) circuit for the two regions A and B in question.

i) Infected individuals in B may infect others staying in B, including those traveling from other regions, as in Eqs. (11)–(14).

ii) Individuals traveling from A to other regions are added to the group of the same state traveling from A as in Eqs. (15)–(18), as well as infect them as in (i).

As a result, the model we developed and used in this study is represented in Eqs. (1) through (4) and illustrated in Fig.2**c**. We set $m_{SER}$ to zero and ignored terms that vanished by this setting in Eqs. (11) through (18).

About the parameter values in Eqs. (1) through (4), we set $\alpha$ to 1 for the most recent normal year, that is, 2019. Assuming a decrease of 60% in year 2021 in comparison with 2019 (on the statistical data [36,37]), $\alpha$ of 0.4, is regarded to approximate the real current status. $T_{kj}$ is obtained as 2.3 $T_1$+ $T_2$, referring to the data [13, 38] for $T_1$, meaning the number of travelers who stay in region $j$ (for 2.3 days as mentioned above) on average and $T_2$ of a one-day trip. Specifically, we estimated the approximate frequencies of movements within Tokyo and Tokyo in an ordinary year before 2020 from [38] and used the results to estimate the corresponding values for other prefectures. Then, these values for each prefecture (e.g., $j$) have been divided into $T_{kj}$ representing the movements from/to each to other prefectures ($k$'s) in proportion to the population of the destinations ($k$'s).

## Data Availability

The data obtained and used for this paper are put in
 https://www.panda.sys.t.u-tokyo.ac.jp/natmet_Ohsawa.zip
All the data used for this study are included in the reference.

## Acknowledgment

This study was supported by the Cabinet Secretariat of Japan, both financially and via scientific communication. We also appreciate the researchers in Trust Architecture Inc. for their insightful communications. In addition, the computational basis of this study has been partially supported by a grant from JSPS Kakenhi 20K20482.

Contributions

Modeling: Ohsawa
Conceptualization: Ohsawa for Lattice of SEIRS Circuits, and Kondo for Social Muddling
Coding: Ohsawa
Data collection: Ohsawa, Hayashi
Interpretation of the results: Ohsawa, Hayashi, Kondo
Writing: Ohsawa, Hayashi, Kondo
Fund acquisition: Ohsawa

## Supplementary Information

**Fig. S1 (hyperlink)** The effect, in the case involving the variant B.1.617, of the conditional entropy $H_c$ (in **a-l**) and the average $R_t$ (in **m-x**) of vaccinated prefectures on the number of infection cases (accumulation), varying the vaccination pace $p_v$ and the travel activity $\alpha$. All the parts of this figure mean the same as Fig.6 in the main text. A qualitative difference from Fig.6 is the positive correlation of the number of infected cases to $R_t$ as in **v, w, x** that was negative in the case of B.1.1.7.